# The Evolution of User-Selected Passwords: A Quantitative Analysis of Publicly Available Datasets

Theodosis Mourouzis (theodosis@ciim.ac.cy)[1,2]
Kyriacos E. Pavlou (kyriacos.pavlou@ciim.ac.cy)[1,2,*]
Stylianos Kampakis (stylianos.kampakis@gmail.com)[3]

[1] Cyprus International Institute of Management, 21 Akademias Ave, Nicosia 2107, Cyprus

[2] UCL Centre for Blockchain Technologies, UCL Computer Science, Malet Place, London, WC1E 6BT, UK.

[3] UCL Computer Science, University College London, Gower St, Bloomsbury, London WC1E 6BT, UK

## Abstract

The aim of this work is to study the evolution of password selection among users. We investigate whether users follow best practices when selecting passwords and identify areas in need of improvement.

Four distinct publicly-available password datasets (obtained from security breaches, compiled by security experts, and designated as containing bad passwords) are employed. As these datasets were released at different times, the distributions characterizing these datasets suggest a chronological evolution of password selection.

A similarity metric, *Levenshtein distance*, is used to compare passwords in each dataset against the designated benchmark of bad passwords. The resulting distributions of normalized similarity scores are then compared to each other. The comparison reveals an overall increase in the mean of the similarity distributions corresponding to more recent datasets, implying a shift away from the use of bad passwords.

This conclusion is corroborated by the passwords' clustering behavior. An encoding capturing best practices maps passwords to a high dimensional space over which a *k*-means clustering (with silhouette coefficient) analysis is performed. Cluster comparison and character frequency analysis indicates an improvement in password selection over time with respect to certain features (length, mixing character types), yet certain discouraged practices (name inclusion, selection bias) still persist.

*Corresponding Author

# 1 Introduction

The design of a secure, usable, and user-friendly authentication scheme is among the top priorities for business and organizations that provide online applications. Authentication refers to techniques applied in order to confirm the validity of an attribute of a single piece of data claimed as true by a specific entity. Data can be authenticated based on several factor categories [10, 23]:

*Knowledge factors*: something the user *knows*, like a password, a partial password [12], a passphrase [17], a personal identification number (PIN), or a security question.

*Ownership factors*: something the user *has*, like a hardware or software token, a cell phone, or an ID card.

*Inherence factors*: something the user *is* or *does*, e.g., fingerprints, retinal patterns, DNA sequences, face characteristics, mouse and keystroke characteristics and other biometrics.

The security research community has determined that for a sufficiently high level of security, at least two factors, each from a distinct category, must be combined in an authentication scheme. This is known as a two-factor *authentication scheme* and it belongs to the family of *Multi-Factor Authentication (MFA) schemes* [10].

MFA schemes are now considered as the de-facto authentication mechanism in the industrial world. Moreover, a significant amount of money, time and effort are invested every year in order to promote research for strengthening the existing MFA techniques in an attempt to increase online security for companies, stores and organizations. According to Cybersecurity Ventures, a leading research company specializing in global cybereconomy, the U.S. Government has invested more than 50 million dollars from 2011 to 2015 in order to promote the Multi-Factor Authentication market in collaboration with research and technology institutions [6]. In general, one trillion dollars will be spent globally on cybersecurity products and services in the period 2017–2021. Similarly, the worldwide cybersecurity market grew from 3.5 billion US dollars in 2014 to 75 billion in 2015, and is forecast to reach 170 billion dollars by 2020 [7].

As a result of these tremendous research efforts, we now have new and potentially stronger methods of authentication, such as biometrics, mouse and keystroke analytics as mentioned previously.

However, passwords, despite being an old technology remain the primary means of authentication for many online services. For example, most banks that now employ MFA schemes for online banking authentication [2], use combinations in their deployed two-factor schemes that almost always include passwords. These common combinations are: a) a password with a hardware token, b) a password with a PIN or c) a password with a partial password [2,12]. Furthermore, recent studies have shown that the public is not satisfied with other authentication techniques such as hardware tokens used in Internet Banking [1]. The persistence and popularity of passwords as an important mechanism of authentication is due to their proven high-usability, ease-of-use, and efficiency regardless of any concerns about their security [1].

**Table 1.** Rules/Policies for Password Selection

| 1. | Use of mixed upper and lowercase letters |
|---|---|
| 2. | Use of at least one numeric digit |
| 3. | Use of special characters such as @, $, # |
| 4. | No use of words found on password blacklists |
| 5. | No use of words found in the user's personal information (e.g., name, date of birth, abbreviations) |
| 6. | No use of passwords that match the format of calendar dates, license plate numbers, telephone numbers, or other common numbers |
| 7. | Avoid use of same passwords across several services |

Unfortunately, password-based authentication techniques present security challenges. A severe drawback of such mechanisms is the fact that their security relies on the ability of human users to follow best practices when choosing passwords. In fact, this could be considered as a single point of failure for this mechanism. Either out of laziness or ignorance, users often reuse passwords or select weak$^\dagger$ passwords that are convenient to remember [22].

The situation is further compounded by the rapid expansion of internet-based services which all require a password for authentication. As a result, each user needs to remember and keep secure a continuously growing list of online usernames and passwords.

According to a survey by managed services provider Redcentric, more than three in five people are putting themselves at risk online by using the same password across multiple online accounts [25]. In addition, the same survey revealed that 21% of people only change their password when prompted, 33% of people's password contain personal information, such as personal names, addresses and birthdays while 17% of people admitted to storing their password on their computer or mobile phone. Thus, for users, password management becomes an increasingly complex issue.

Even if people do their best to follow best practices for password selection, they turn out to be vulnerable to social engineering attacks, under which they are successfully manipulated by malicious parties into revealing their credentials [23].
The academic and industrial communities, recognizing the importance of password-based schemes and in an attempt to mitigate the dangers inherent in them, have encouraged the use of strong passwords by developing a set of password selection Information Security (IS) policies. These policies summarized in Table 1, indicate several rules to follow during the selection and usage of a password [28].

The aim of this paper is to study whether security, education, training, and awareness (SETA) programs as well as exposure to security incidences over the years have influenced user behavior as reflected in the evolution over time of password selection. (In this paper, the term "password selection" refers to the act of individuals/end users choosing specific alphanumeric characters and special symbols and concatenating them in order to create a password.) Specifically, we would like to ascertain whether users are becoming better at following policies and thus selecting stronger passwords. Such a longitudinal study requires studying the behavior of human subjects over time. This presents many challenges not only because people are reluctant to reveal their passwords, the study has to be performed over a large population over a long time.

---

$^\dagger$The standard definition of password strength, formulated in terms of information entropy, applies.

In order to achieve our aim and avoid the complications of a longitudinal study we use a different approach: using known statistical analysis techniques we study the underlying distributions and similarities of several user-selected passwords based on leaked or compiled real-world password datasets. In particular, we have used the following datasets: `MySpace` [4], `phpBB` [4], and `RockYou` [4] and the more recent `Xato` dataset [5]. Running several experiments on these large datasets allows us to tease out changes in user behavior as reflected by password selection.

The rest of the paper is organized as follows: Section 2 presents studies related to password security and the distribution of user-selected passwords. In Section 3, we study the similarity of the four datasets compared to a set of bad passwords that do not follow best security practices. In Section 4, we perform a clustering analysis on these four datasets, by encoding the underlying features according to best security practices. Section 5 studies the distribution of password lengths as well as the frequency distributions of digits and special characters appearing in the datasets. Lastly, in Section 6 we summarize our results and conclude with further research plans.

## 2 Related Work

There has been a great deal of work studying different aspects of passwords and evaluating IS policies and user behavior.

Kelley et al. studied different metrics used in characterizing text-based passwords and used these metrics in order to evaluate password-selection policies [18]. They introduced a technique for password strength evaluation that can be implemented for a range of password-guessing algorithms and tuned these algorithms using a variety of training sets to gain insight into the comparative guess resistance of different sets of passwords. Using this methodology, they compared the strength of different composition policies.

Jakobsson et al. studied the generation of passwords selected by users in an attempt to distinguish good passwords from bad ones [16]. They revealed that users produce passwords by using a much smaller subset of rules and types of components compared to the rules imposed by password policies.

User behavior in password selection were studied and various authentication approaches were presented in a comprehensive paper by Yampolskiy [29]. After analysis of several user authentication mechanisms the author concluded that any authentication system, which allows involvement of users in the password selection process is vulnerable to password space reduction methods and should thus be avoided.

Egelman et al. conducted lab experiments to examine the effect of password meters, which inform users about password strength, on the password selection process [9]. The authors concluded that the presence of meters yielded significantly stronger passwords. However, a follow-up experiment showed that such meters only have a significant impact when forcing users to change their password for important accounts while for low-risk accounts the presence of such meters yielded no observable difference.

In a 1992 article Spafford identified that the common method for addressing the issue of weak passwords compromising the security of computer systems is problematic [27]. Specifically, the efficiency of comparing user-selected passwords against a list of unacceptable words is highly-dependent on the size of said list. The author described a space-efficient method of storing a dictionary of words that are not allowed as password choices. Lookups in the dictionary take constant time, regardless of dictionary size.

Yan et al. conducted controlled trials and managed to confirm a number of widely held folk beliefs about passwords while debunking others [30]. They confirmed that users have difficulty remembering random passwords and that passwords based on mnemonic phrases are harder to guess than naively selected ones. The authors debunked the belief that random passwords are better than passwords based on mnemonic phrases: both appear equally strong. Another belief likewise debunked is that passwords based on mnemonic phrases are harder to remember than naively selected passwords: both appear to be equally memorable.

In a 2014 article Bonneau and Schechter challenged the conventional wisdom that users cannot remember cryptographically-strong secrets [3]. The authors tested successfully the hypothesis that users can learn randomly-assigned 56-bit codes (encoded as either 6 words or 12 characters) through spaced repetition. They debunked the myth that users are inherently incapable of remembering cryptographically-strong secrets for a select few high-stakes scenarios.

In addition to the study of password policies, the effect of such IS policies on employees and users, as well as the methods for achieving better policy compliance were extensively studied [8, 14, 19, 21].

Hsu et al. examined the importance of in-role (security behaviors specified in IS policies) and extra-role behaviors (security behaviors that benefit organizations but are not specified in IS policies) in the effectiveness of said policies [14]. The authors collected data from 78 IS managers and 217 employees from the same organization and concluded, based on social control theory, that both formal and social controls enhance in-role and extra-role behaviors. This work suggests that extra-role security behaviors should be encouraged in order to improve security and enhance intra-organization employee connections.

In a survey conducted on 186 employees, it was found that punishment expectancy is a strong determinant of compliance behavior, whereas the main effect of reward expectancy is not significant [21]. This suggests that employee compliance behavior is heavily affected by the disciplinary actions related to non-compliance. Another study on 269 computer users from eight different companies, empirically showed that IS misuse is directly related to awareness of security policies, security education, training, SETA programs and computer [8].

Compliance is crucial in information security, as industry statistics suggest that 50% to 75% of security incidents originate from within an organization [8]. That is a significantly large percentage that we need to understand how to reduce. Precisely for this reason, in the series of ISO 27001 standards pertaining to IS a lot of emphasis is given to Human Resource (HR) security, user access management, and user training, awareness, and compliance [15].

Kirlappos et al. identified (in a previous study) a new category of employee security behavior: *shadow security*. This behavior comprised methods employees devise to compromise between success in primary business goals (i.e., "getting the work done") and extant IS policies. The authors discussed new findings from an interview study they conducted in multinational organizations [19]. Even though shadow security behavior may not be compliant with official policy or may be downright risky, it should not be dismissed as a problem. Once identified, such conduct should be treated as an opportunity to identify inadequacies within current IS policy implementations and as a guide in devising solutions to address said inadequacies. This enables an environment in which organizations continuously revise and improve their IS practices.

Recently, Microsoft has enhanced its password policy for Active Directory and Azure Active Directory with more rules. These changes were deemed necessary after analyzing millions of username and password pairs received every day [13]. That these platforms receive over 10 million such pairs on a daily basis implies that this huge amount of data can be used to approximate the entire population's password selection behavior. Apart from reiterating known policies such as the 8-character minimum password length requirement, password-reuse avoidance, and multi-factor authentication, Microsoft's report suggests policies that are not consistent with certain established password selection policies. One suggestion is the elimination of mandatory periodic password resets for user accounts because the new password chosen is easy to predict from the old password. The most interesting, however, is the advice to IT administrators to eliminate character-composition requirements, i.e., the mixing of uppercase, lowercase, numerical, and special character sets. The reason given is that most people when creating a password use similar patterns (e.g., capital letter in the first position, a symbol in the last, and a number in the last two positions) thus making passwords more predictable. This is an active area of research and further study is needed to decide definitively the merit of the above recommendation[‡].

The current paper carries several contributions. It significantly extends previous work on user password selection behavior by analyzing real-world passwords while not relying on subjective approaches like self-reporting. The datasets involved in the analysis are quite large (`MySpace.txt`: 356KB, `phpBB.txt`: 1.6MB, `RockYou.txt`: 139.9MB and `Xato.txt`: 95.9MB) leading to a more realistic description of the evolution of password selection. Finally, our analysis reveals which IS policies for password selection are successfully employed by users and which are not.

## 3 Similarities in User-Selected Password Datasets

In order to test our hypothesis, whether the user-selected password distribution shifts towards stronger passwords or not, we have computed the similarity of four distinct password datasets released at different chronological orders, spanning almost a decade, against a large dataset containing known bad passwords.

After we compute the similarity scores between the datasets, we perform unpaired (two sample) $t$–tests in order to prove or disprove any statistical significance of the differences observed. The datasets used in this experimental work are public. They were released for research purposes and were obtained either from several different online sites via attacks such as SQL injections, malware and phishing attacks [4], or they were compiled by information security experts [5]. We assume that these datasets reflect, to a very high degree, the way users select passwords and that they capture sufficient structure from the general distribution of user-selected passwords.

---

[‡]This paper follows the recommendations of industry standards on character composition as outlined in ISO27001.

**Figure 1.** Testing & Analysis Setup

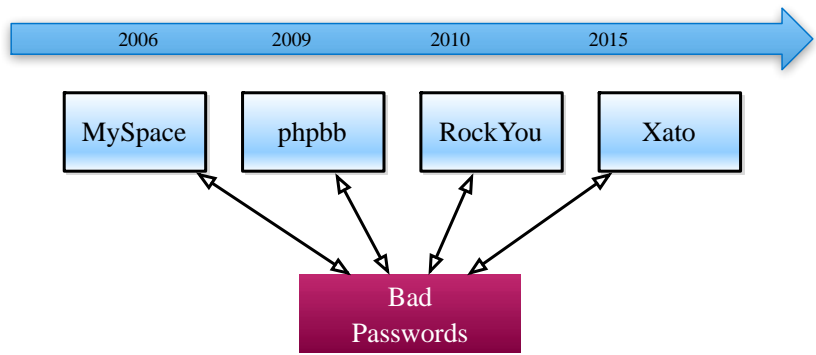

The following datasets, available on the SkullSecurity Blog, were used [4].

1. `MySpace`: This dataset is claimed to have been obtained via phishing techniques [4]. It contains approximately 37,000 passwords selected by users of the online social network MySpace. It was released around October 2006.

2. `phpBB`: This dataset is claimed to have been obtained by cracking the MD5 hashes contained in the password file obtained after an attack, possibly an SQL injection attack, on the phpBB website [4]. This dataset dates back to early 2009 and contains approximately 184,000 passwords.

3. `RockYou`: This is the largest and most comprehensive password list originating from a single source available. It is also the most and well-studied of all such datasets. This dataset was obtained by an attack on the RockYou company database, a company that develops widgets for MySpace and implements applications for various social networks like Facebook. The password list was stolen in an unencrypted form. The list contains approximately 14,000,000 distinct passwords and dates back to 2010.

4. `Xato`: This is a dataset compiled and released by Mark Burnett, an IT security analyst, on his personal blog [5]. It is a collection of passwords that were collected over the last fifteen years and comprises the combined data samples obtained from thousands of global incidents. It was released in February of 2015 and contains around 10,000,000 user-selected passwords.

5. `Bad Passwords`: This file was compiled for the purposes of this work. In this list, we have included passwords that were either banned from several companies and organizations or were classified as bad after several studies [4]. In particular, we have combined the "500 worst passwords" and "370 Banned Twitter" datasets found on [4]. This list is made up of approximately 1000 passwords.

Table 2. Statistics reported for each password dataset.

| Statistic | Datasets | | | |
|---|---|---|---|---|
| | MySpace | phpBB | RockYou | Xato |
| Min | 0.0 | 0.0 | 0.0 | 0.1 |
| Max | 1.0 | 1.0 | 1.0 | 1.0 |
| Mean | 0.177 | 0.235 | 0.325 | 0.212 |
| Standard Deviation | 0.162 | 0.115 | 0.113 | 0.131 |
| Skewness | 0.991 | 0.548 | 0.580 | 1.820 |
| 1st Quartile | 0.0 | 0.143 | 0.250 | 0.143 |
| Median | 0.143 | 0.222 | 0.333 | 0.143 |
| 3rd Quartile | 0.286 | 0.333 | 0.400 | 0.25 |

The datasets MySpace, phpBB and RockYou span a period of approximately five years, while Xato dataset is a comprehensive dataset that includes data over the last fifteen years. Xato could be seen as an aggregation of user-selected password datasets that represent different clusters of users for a longer period.

## 3.1 Similarity Metrics

The similarity tests among password files use the *Levenshtein* distance (or edit distance) [20].

**Definition 3.1.** Given two strings $s_1$, $s_2$, of length $|s_1|$, $|s_1|$ respectively, then the *Levenshtein distance*, denoted by $lev_{s_1,s_2}(|s_1|,|s_2|)$ is given by:

$$lev_{s_1,s_2}(i,j) = \begin{cases} \max(i,j), & \text{if } \min(i,j) = 0 \\ \min \begin{cases} lev_{s_1,s_2}(i-1,j) + 1 \\ lev_{s_1,s_2}(i,j-1) - 1, \\ lev_{s_1,s_2}(i-1,j-1) + I_{s_{1_i} \neq s_{2_j}} \end{cases} & \text{otherwise} \end{cases}$$

where $I_{s_{1_i} \neq s_{2_j}}$ is an indicator function equal to 0 when $s_{1_i} = s_{2_j}$ and equal to 1 otherwise, while $lev_{s_1,s_2}(i,j)$ is the distance between the first $i$ characters of $s_1$ and the first $j$ characters of $s_2$.

The edit distance between two words represents the minimum number of single-character edits, that is, insertions, deletions, substitutions, required to transform one word into the other [20]. For our experimental simulations, we have used the `editdistance 0.3.1` Python package, a fast implementation of the edit distance algorithm written in C++ and Cython [24].

**Figure 2.** The distribution of Levenshtein distances of passwords from each of the four datashets. compared to the benchmark dataset of bad passwords.

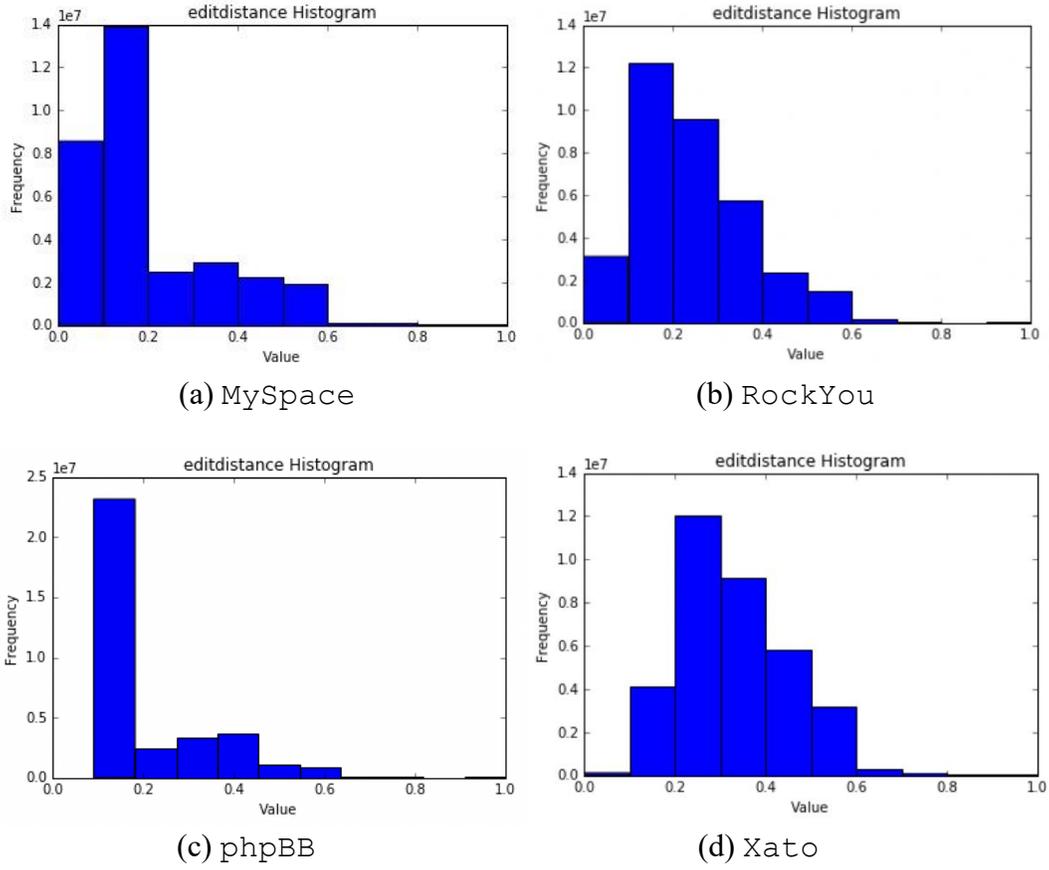

(a) MySpace
(b) RockYou
(c) phpBB
(d) Xato

For the purpose of our experiments we are interested in the distribution of the edit distances of the passwords of a given dataset compared to the benchmarked dataset of bad passwords.

**Definition 3.2.** We define the distance $d_L$ of a password $p_1$ that belongs to dataset $D_1$ from dataset $D_2$, as the minimum distance across all possible edit distances of $p_1$ to each password in $D_2$, given as follows:

$$d_L(p_1, D_2) = \min_{p_2 \in D_2} lev_{p_1, p_2}(|p_1|, |p_2|)$$

Then, we can define the average distance, based on the edit distance metric, between two datasets.

**Definition 3.3.** The average edit distance $\overline{d_L}$ between two datasets $D_1, D_2$ of sizes $|D_1|, |D_2|$ respectively, is defined as follows:

$$\overline{d_L}(D_1, D_2) = \frac{\sum_{p_1 \in D_1} d_L(p_1, D_2)}{|D_1|}$$

## 3.2 Testing for Similarity to Bad Passwords

Datasets released more recently are assumed to represent users that have had the opportunity to be exposed more to security awareness campaigns and were witness to more information security-related incidents. Hence, such exposure could have had an effect on the way users select passwords. Under this assumption, we expect the more recent the datasets are, the less similar they are compared to the bad passwords dataset. In order to test this assumption, we compute the similarity of each of these files compared to a large list of bad passwords (Figure 1). In mathematical terms, we are interested in finding the average similarity between a file $D_1 \in$ {MySpace, phpBB, RockYou, Xato} and $D_2 \equiv$ Bad Passwords.

In order to speed up our simulations, instead of comparing the full dataset $D_1$ to $D_2$, we have repeatedly used samples of size 40,000. Since the sample is sufficiently large, we can apply the Central Limit Theorem (CLT), in order to estimate confidence intervals for the mean similarity between $D_1$ to $D_2$. Summary results of simulations for samples extracted from $D_1$ are presented in Table 2.

In Table 2 we observe that the distributions of Xato and MySpace are highly positively-skewed, which implies the percentage of passwords with a high similarity distance is lower than the ones with lower similarity distance. In the other two datasets, we have a skewness coefficient close to 0.5 which implies the distributions are fairly symmetrical.

Figures 2a–2d present the distributions of edit distances obtained after computing the individual distances of each password from samples extracted from MySpace, phpBB, RockYou, and Xato respectively compared to the benchmarked dataset of bad passwords.

In Figure 2a we observe that a great proportion of user-selected passwords belonging to the older dataset MySpace are very similar to the ones that are in the dataset of bad passwords. Thus, users of this period were selecting some passwords which are considered insecure. Users of this period might have been exposed less to security events or have been less security-aware. However, according to Figures 2b and 2c, that present the edit-distance distribution for phpBB and RockYou users respectively, we observe that the percentage of passwords that have an edit distance less than 0.1 are greatly reduced compared to those in Figure 2a (compare first bar of each histogram). In addition, the height of other classes that represent passwords with higher edit distance is significantly increased, implying a shift to stronger passwords. We might say that users became more security aware and tech-savvy. In addition, users might have experienced more security-related incidents and thus they turn into selecting more secure passwords.

Xato's distribution, as illustrated in Figure 2d, allows three important observations to be made. One observation is about having no passwords of edit distance less than 0.1—translated to a similarity greater than 90%. This might be interpreted as users applying some tricks in an effort to make the passwords more secure, e.g., simple substitutions of letters with digits. The second observation is about the large class of passwords with edit distance in the range [0.1, 0.2]—translated to 80-90% similarity to bad passwords. We believe that both these classes represent passwords selected during the earlier part of the chronological period the dataset was collected when users where not closely adhering to password selection policies. Lastly, passwords in the rest of the classes might represent more recently selected passwords.

**Figure 3.** The boxplots of all password datasets.

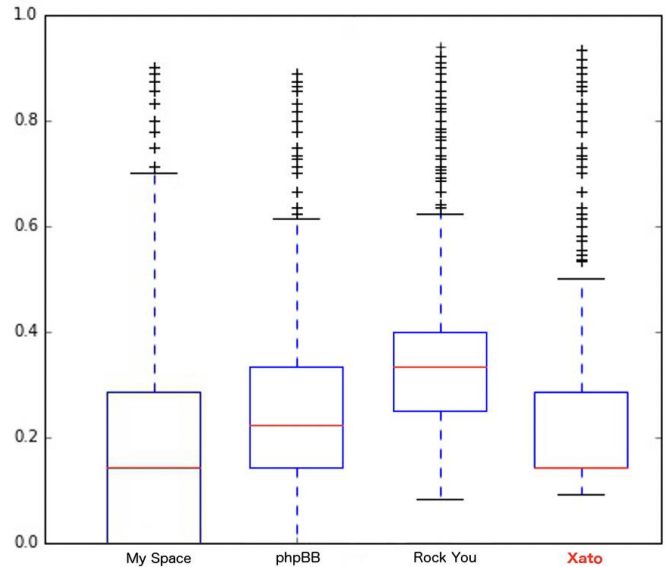

Based on our simulations we observe that users are shifting towards more secure password selections and the user-selected passwords distribution tends to move away from password selections that are considered insecure. In order to visualize this shift in a better way we plotted the boxplots of these distributions in Figure 3. We observe a tendency of the mean edit distance to increase with the passing of time. This is not in the same of the Xato dataset as it contains data compiled over the whole decade. It represents an overall average and acts a sanity check because its average edit distance and boxplot fall within those of the other three datasets.

In order to test the statistical significance of our results we have an unpaired $t$–test to check if there is any statistically significant difference among passwords in MySpace, phpBB and RockYou.

We have tested, the following hypothesis, at a 5% significance level:

$H_0$: No difference between the average similarity distance of datasets $D_1$, $D_2$

$H_1$: There is difference.

We have used an unpaired two-sample $t$-test with unequal variances. For the datasets MySpace and phpBB we have obtained a statistic t = -58.38 ≪ 1.96 (critical point). Thus, based on this result the difference is considered to be extremely statistically significant. For phpBB and RockYou the statistic is approximately t = 111.65 ≫ 1.96 and thus again there is extremely statistical significance between the two means.

**Table 3.** Vectorization Map

| Dimension | Mapping |
|---|---|
| $x_1$ | is a variable defined to encode the length $|p|$ of a password $p$ |
| $x_2$ | equals to 1 if a password comprises both upper- and lowercase letters; 0 otherwise |
| $x_3$ | is the number of lowercase letters |
| $x_4$ | is the number of uppercase letters |
| $x_5$ | equals to 1 if the password is alphanumeric and 0 otherwise |
| $x_6$ | is the number of digits that appear in the password |
| $x_7$ | equals to 1 if the password contains special characters; 0 otherwise |
| $x_8$ | is number of special characters that appear |
| $x_9$ | is the minimum Levenshtein distance from a dataset of bad passwords |
| $x_{10}$ | equals to 1 if it contains a personal name; 0 otherwise |

# 4 Clustering Analysis

In the previous section, we showed that there is a significant statistical difference among the means of the relative edit distance distributions of the four datasets, compared to a set of bad passwords.

Assuming, that these datasets are chronologically ordered we can claim that there is a significant change in the user-selected passwords towards more secure selections, moving away from the bad passwords. In particular, in `Xato`, whose data spans the last fifteen years, we observe a huge percentage of the passwords having a similarity between 0.1 and 0.2 compared to the set of bad passwords, while the rest of the passwords are distributed to other classes quite dissimilar to bad passwords. This could suggest the existence of two clusters of passwords, with one cluster representing weaker passwords selected early on in the fifteen-year period, while the other representing a subsequent move towards more secure selections.

Such observations hint at an underlying structure in these datasets which could help us extract more meaningful conclusion. In this section, we analyze the clustering behavior of the passwords. This would allow us to detect and study the characteristics of clusters comprising passwords that are less secure, i.e., don't comply with best practices in password selection.

In order to study this problem, we employed a standard *vectorization process* whereby each password is mapped to a 10-dimensional space constructed in such a way that each coordinate/feature quantifies compliance with the rules of well-known password policies—enriched with additional policies. Hence, the formation of clusters reveals passwords selected in a similar way. Table 3 presents the map in detail. Please note that all coordinates with the exception of password length ($x_1$) codify in various way the first six of the best policies for password selection given in Table 1. The seventh policy, namely, password reuse, will be discussed in Section 5. In $x_{10}$ of the map we check if the passwords contain personal names by comparing them against a publicly available database comprising circa 5500 (European, Latino, and Asian) first names [11].

The clustering analysis uses the $k$-means algorithm with silhouette coefficient, based on the Euclidean distance metric [26]. The silhouette coefficient is a measure of how similar the objects are within their own cluster compared to other clusters. It is, therefore, a measure of how separable the clusters are. It ranges within [-1, 1], where values close to +1 indicate that the sample is well matched to its own cluster, 0 indicates that the sample is on or very close to the decision boundary between two neighboring clusters, while values close to -1 indicate that samples might have been assigned to the wrong cluster.

Table 4. The clusters and their characteristics for each password dataset.

| Dataset | Silhouette Coefficient | Clusters | Description |
|---|---|---|---|
| MySpace | 0.81 | myspace_1: [ 7.13, 0. , 4.65, 0.17, 1. , 2.88, 0. , 0.04, 0.61, 0. ]<br>myspace_2: [ 8.58, 0. , 6.29, 0.21, 1. , 1.84, 0. , 0.03, 0.57, 1. ]<br>myspace_3: [ 8.77, 0.04, 7.85, 0.46, 0. , 0. , 1. , 1.81, 0.54, 1. ]<br>myspace_4: [11.07, 0.03, 5.65, 0.3 , 0. , 0. , 0.82, 1. , 5.92, 0.58, 0. ]<br>myspace_5: [ 8.28, 1. , 4.75, 1.37, 1. , 2. , 0. , 0.14, 0.62, 0.66 ] | · length of password not an identifying factor<br>· no mix of lower and upper alphabet letters<br>· mainly use of lower case letters<br>· almost no digits or special characters used<br>· three clusters contain personal names |
| phpBB | 0.63 | phpbb_1: [ 6.5 , 0. , 4.16, 0.14, 1. , 3.13, 0. , 0.03, 0.65, 0. ]<br>phpbb_2: [ 6.75, 0.05, 7.5 , 0.22, 0. , 0. , 0. , 0.02, 0.54, 1. ]<br>phpbb_3: [ 4.17, 0. , 3.41, 0.08, 0. , 3.22, 0. , 0.04, 0.59, 0. ]<br>phpbb_4: [ 8.16, 0. , 5.73, 0.1 , 1. , 2.3 , 0. , 0.02, 0.59, 1. ]<br>phpbb_5: [ 8.07, 1. , 3.95, 2.16, 0.83, 1.7 , 0. , 0.06, 0.7 , 0.3 ] | · length is an identifying factor<br>· no mix of lower and upper characters<br>· mainly lower case letters<br>· no class make use of special characters and only a few use digits<br>· three clusters contain personal names |
| RockYou | 0.72 | rockyou_1: [ 7.43, 0.02, 1.82, 0.15, 0. , 6. , 0. , 0.11, 0.64, 0. ]<br>rockyou_2: [ 9.78, 0. , 6.19, 0.36, 1. , 2.64, 0. , 0.09, 0.63, 1. ]<br>rockyou_3: [ 9.13, 0. , 8.56, 0.48, 0. , 0. , 0. , 0.13, 0.6 , 1. ]<br>rockyou_4: [ 9.4 , 1. , 5.41, 1.65, 0.74, 1.87, 0. , 0.22, 0.65, 0.78 ]<br>rockyou_5: [ 7.55, 0. , 4.12, 0.23, 1. , 3.6 , 0. , 0.09, 0.66, 0. ] | · length is not an identifying factor but longer passwords<br>· no mix of lower and upper characters<br>· half of the clusters use on average two digits<br>· most passwords do not use special characters<br>· half of the clusters contain personal names |
| Xato | 0.68 | xato_1: [ 7.98, 0. , 2.29, 0.09, 1. , 4.52, 1. , 1.01, 0.62, 0. ]<br>xato_2: [ 8.63, 0. , 7.2 , 0.14, 0. , 0. , 1. , 1.01, 0.54, 1. ]<br>xato_3: [10.6 , 0. , 5.94, 0.11, 1. , 2.63, 1. , 1.02, 0.63, 1. ]<br>xato_4: [ 9.69, 0. , 3.98, 0.15, 1. , 3.83, 1. , 1.01, 0.68, 0. ]<br>xato_5: [ 9.75, 1. , 3.61, 2.63, 0.78, 1.89, 1. , 1.03, 0.75, 0. ]<br>xato_6: [10.4 , 1. , 5.41, 1.95, 0.6 , 1.26, 1. , 1.04, 0.66, 1. ] | · passwords vary from 8 to 10<br>· two clusters mix lower and upper case letters<br>· most of the clusters use on average 2 digits<br>· all clusters contain at least one special character<br>· half of the clusters contain personal names |

The vectorization results for each cluster centroid are tabulated in Table 4. The silhouette coefficients have sufficiently high values for all datasets (Silhouette Coefficient ≥ 0.63), implying that cluster elements are very similar within their cluster. For datasets `MySpace`, `phpBB`, `RockYou` we have obtained five clusters each, while for `Xato` dataset we have obtained six clusters.

Analysis has shown the existence of five clusters within the `MySpace` dataset. In four out of the five clusters, `myspace_{1-3,5}`, there are passwords of average length 8 and one cluster, `myspace_4` that contains longer passwords of average length 11. Mixing of lowercase and uppercase alphabet letters is not seen in the majority of clusters, except in `myspace_5`. However, we have the majority of clusters with passwords using digits, except `myspace_{3,4}`. Special characters appear only in clusters `myspace_{3,4}` using 1.81 and 5.92 special digits respectively. The Levenshtein distance is not an identifying factor between the clusters with the values being comparable to those in other dataset clusters. In three out of the five clusters, namely `myspace_{2,3,5}`, we have passwords that contain personal names, making the passwords in these clusters quite insecure. The `myspace_4` cluster is closer to the best policy standards if you ignore the fact that mixing of lowercase and uppercase alphabet letters is rarely observed within the cluster.

The `phpBB` dataset features five clusters. Password length is not an identifying factor between the clusters, with the sole exception of `phpbb_3` in which passwords have the smallest average length 4.17. Mixing of both uppercase and lowercase alphabet characters is still not observed by the majority of clusters, except in cluster `phpbb_5`. Most of the clusters contain passwords that include, on average, more digits than the clusters in the `MySpace` dataset but in none of the clusters the use of special characters is an identifying factor. However, the majority of the clusters (three out of five), have passwords that include personal names. The Levenshtein distance of the passwords in the `phpBB` clusters is higher, on average, than those in the `MySpace` clusters.

Analysis of the `RockYou` dataset has also yielded five distinct clusters. The length of the password is not an identifying factor across the clusters but it appears to, overall, be higher compared to previous datasets. We observe mixing of lowercase and uppercase alphabet characters in only one cluster out of five, namely, `rockyou_4`. The use of digits is seen in three out of the five clusters. However, passwords in `rockyou_1` seem to have few letters (lowercase $x_3$= 1.82, uppercase $x_4$= 0.15) suggesting that they consist mainly of digits. This is highly problematic, as it might indicate the existence of telephone numbers, social security numbers or any other user-identifying information being as passwords. The use of special characters is virtually non-existent ($x_7$= 0.0) in the passwords clusters of this dataset. The average Levenshtein distance ($\bar{x}_9$) is higher compared to those in the previous datasets, revealing a gradual improvement in password quality. Despite this, there are still three clusters,

**Table 5.** Statistics for each dataset regarding the distribution of password length.

| Password Length ($l$) | Percentage of Passwords of Length $l$ | | | |
|---|---|---|---|---|
| | MySpace | phpbb | RockYou | Xato |
| 6 | 15.2 | 22.8 | 13.4 | 5.0 |
| 7 | 22.6 | 17.8 | 17.5 | 25.4 |
| 8 | **22.9** | **30.0** | **20.7** | 16.6 |
| 9 | 17.3 | 10.4 | 15.3 | **30.0** |
| 10 | 14.3 | 6.5 | 14.0 | 6.8 |
| Total % for $8 \leq l \leq 24$ | 60.5 | 66.9 | 52.0 | 66.1 |

`rockyou_{2-4}`, whose passwords include personal names.

The largest dataset, `Xato`, features a total of six clusters. Password length, which ranges between 7.98 and 10.4 can used to differentiate between the clusters. In terms of mixing of upper and lowercase letters, this is observed in only two clusters: `xato_{5,6}`. Digits are used in four out of the six clusters and, surprisingly, we also observe the use of at least one special character in all clusters. However, we still have passwords that contain personal names in clusters `xato_{2,3,6}`.

In summary, users seem to have learned how to employ some best security practices when choosing a password, but fail to follow all of them. Even though we observe more and more users selecting longer passwords, mixing letters with (usually more than two) digits, and not using passwords similar to known bad passwords, we still rarely see applications of some of the suggested security principles such as mixing of lower with upper case letters, or use of special characters. In addition, passwords that contain special characters are primarily using a single special character.

# 5 Distribution of Password Length, Digits & Special Characters

In this section, we study the distributions of password length, digit frequency and special character frequency in more detail.

## 5.1 Distribution of Password Length

Conventional experience suggests that when users select a secret piece of information that could easily be memorized they might select shorter passwords than required. In order to investigate the veracity of the above statement we turn to a more detailed study of the distribution of password lengths. Specifically, the distribution of password length is computed for all four datasets. The results obtained by our experiments are presented in Figure 4 and summarized in Table 5.

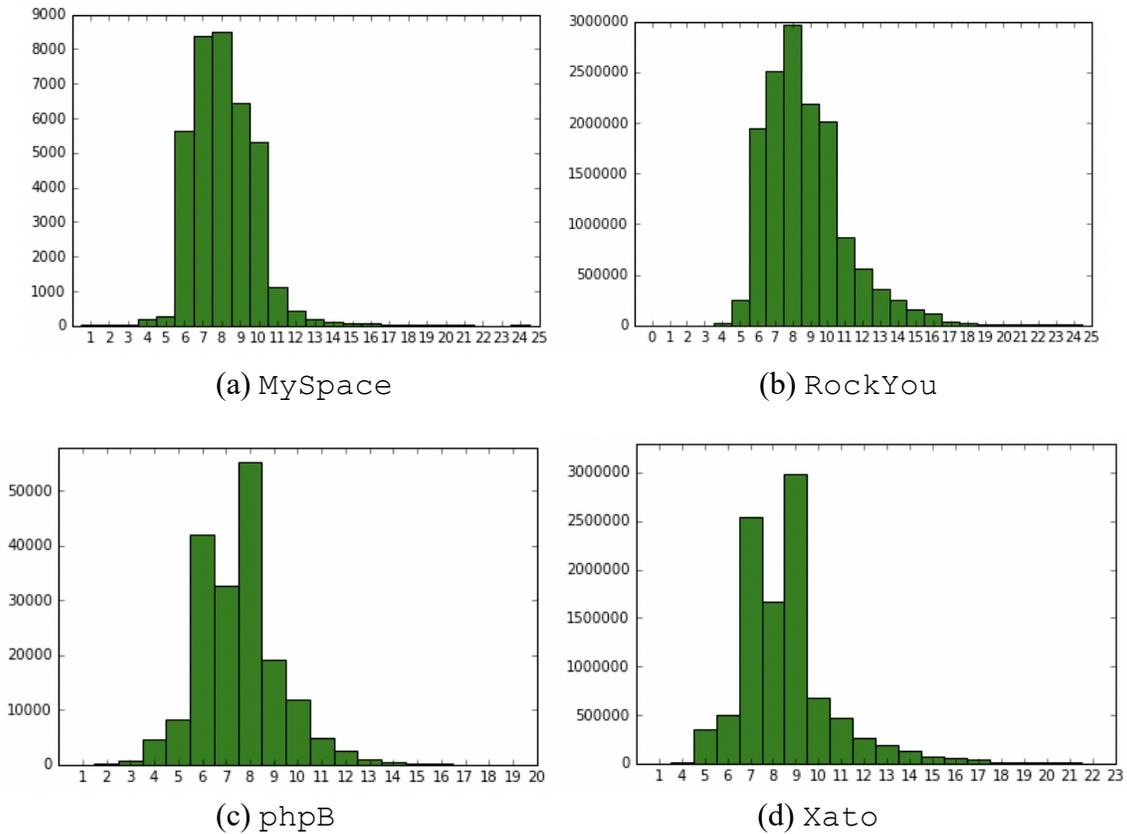

**Figure 4.** The distribution of password lengths in the four datasets.
(Password length vs Number of Passwords)

(a) MySpace  (b) RockYou
(c) phpB  (d) Xato

We observe from Figure 4 that the overwhelming majority of users select passwords of length between 6 and 10. The mode for the first three datasets is length of size 8 with specific percentages shown in Table 5 for MySpace (22.9%), phpBB (30.0%) and RockYou (20.7%). In Xato we observe that passwords of length 9 are the most frequent, with approximately 30.0% of the population having this length. This is particularly encouraging because most password policies suggest selecting a password of at least length 8. Moreover, in all datasets more than half—in some cases more than two thirds—of all passwords have length exceeding the recommended length. We can safely conclude that conventional experience is misleading and that password policies regarding the suggested password length has made an impact on the users' password selection behavior.

## 5.2 Distribution of Digits

In Section 4 clustering analysis revealed that users include at least one digit in their password. In this section, we investigate if users are biased towards selecting specific digits and especially digits that resemble letters.

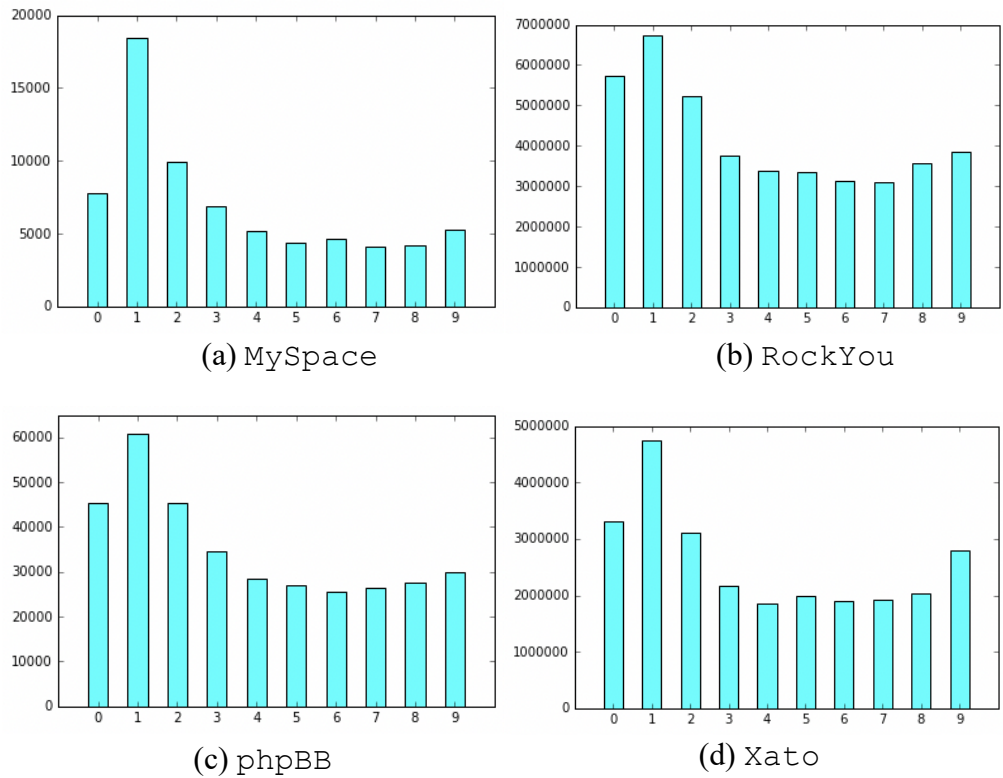

**Figure 5.** The distribution of digits [0–9] used in passwords in the four datasets. (Digits vs Number of Passwords)

(a) `MySpace`  (b) `RockYou`
(c) `phpBB`    (d) `Xato`

Figure 5 illustrates the results obtained from plotting the distribution of digits for all datasets. We observe that in all cases digit 1 is always the most frequently selected digit. Moreover, across all samples digits 0 and 2 and 9 are also very popular. In fact, the exact same pattern exists in all datasets: a distribution in which the most frequently-picked digits are the ones with the lowest and highest values. Even though the bar charts depict the numbers in increasing order because arithmetically 0 comes before 1, keep in mind that on a keyboard the 0 key is located to the right of the 9 key. Thus, the distributions reveal a clear selection bias for digits located at the two extremes of the numerical keys row. This non-uniform preference in digits is problematic since it renders passwords vulnerable to attacks specifically designed to exploit this. For example, in a brute-force attack in which a program starts computing all combinations of digits and letters in order to output candidates for passwords, known patterns such these can speed up tremendously the discovery of passwords that are more likely to be valid passwords. Continuous monitoring and analysis is required of this phenomenon with a concomitant update to password selection policies specifically highlighting the dangers of selection bias.

## 5.3 Distribution of Special Characters

We have already seen that various dataset clusters have passwords that contain special characters (Table 4). The majority of user-selected passwords studied include at least, and in most cases, at

**Figure 6.** The distribution of special characters used in passwords in the four datasets.

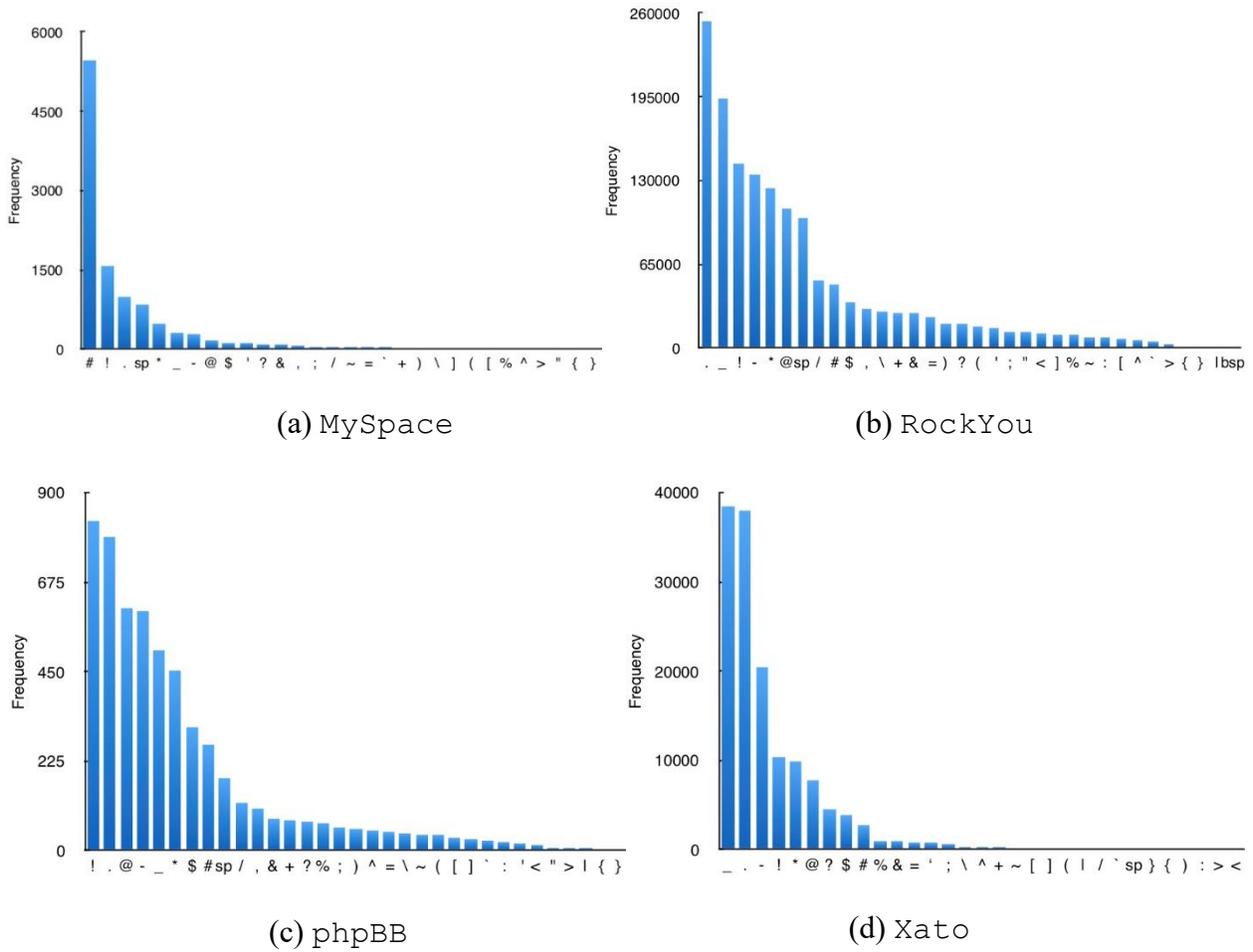

(a) MySpace

(b) RockYou

(c) phpBB

(d) Xato

most one special character. An interesting question that arises immediately from this observation is whether there is any selection bias towards specific characters. In order to address this question, we compute the frequency distributions of special characters for all datasets. The results are illustrated in Figure 6.

Figure 6a shows that in MySpace, the oldest dataset, the most frequent characters are !, #, ., and [space] (abbr. sp), with character # being by far the most widely selected among all. Thus, in this case, the inclusion of a special character in a password does not add the expected value against brute-force attacks as explained previously in Section 5.2. Careful study of the other three datasets, RockYou, phpBB, and Xato, reveals that the six most popular special characters are the same albeit with different order for each dataset. These six special characters are !, ., @, -, _, *. These special characters are frequently used punctuation marks as well as characters used in filenames or email addresses. Again, we observe a selection bias regarding the inclusion of special characters in passwords.

Another remarkable observation that we have made studying the datasets is that a significant number of users use as passwords email addresses from different accounts/providers. We have verified that this occurs in three datasets, namely, MySpace, RockYou, and phpBB.

**Table 6.** Email addresses used as passwords for each dataset.
(The email addresses have been anonymized due to privacy concerns.)

| Data Sample | Email Address Examples | Number of Email Addresses |
|---|---|---|
| MySpace | *i**yh*m*h*e*****1*@hotmail.co.uk <br> *a*t*r**ag*ci***b**t*@gmail.com <br> ***@aol.com | 17 |
| Rock You | *23******@qq.com <br> *7*4**@yahoo.com <br> *s***a@cox.net | ~20,000 |
| phpBB | ***@volga.fr <br> d**z*@mail.ru <br> a*t*o**i*@yahoo.com | 47 |

Table 6 provides some examples of email addresses that were used as passwords in the analyzed datasets. Note that the `RockYou` dataset has around 20,000 email addresses! This is a novel finding which requires further investigation. It is very important that existing password policies make clear that users must never use email addresses as password credentials.

# 6 Conclusion

The security of any type of user authentication method is a topic of major concern for most organizations and enterprises, as it is potentially the primary web gateway to their assets. Precisely for this reason, huge resources are invested every year for the improvement of the security level existing authentication mechanisms offer, as well as the development of new, potentially more secure, ones. As a result of this effort, many different types of user authentication mechanisms exist currently, such as hardware tokens, biometrics, mouse and keystroke characteristics and many others.

The design of a secure authentication scheme is often complicated by the attempt to take into account usability and user-friendliness. Companies and organizations would like their customers to feel that they interact with a user-friendly environment without being overburdened with complex authentication procedures. Therefore, even though, potentially more secure authentication techniques exist, a simple password-based authentication scheme is still the primary means of authentication for many online services.

Password-based authentication techniques have been studied extensively by both the industrial and academic community and several drawbacks have been identified. One of the major drawbacks of this mechanism is that it relies on the assumption that users select sufficiently strong passwords. However, in an effort to select a password they can memorize, users tend to select passwords that are not random at all. This results in selections that might include publicly available information such as personal names, addresses or any other information that can make the password less secure against dictionary attacks. In an effort to ensure users are selecting more secure passwords, password policies comprising of rules regarding how a password should be selected have been developed.

In this paper, we have studied the evolution of user-selected passwords, in order to understand if users exposed to more security related events have become more prudent when selecting their passwords. We have used publicly available datasets, that span over a decade, all containing user-selected passwords. These datasets were leaked from popular websites, such as `MySpace`, `phpBB` and `RockYou` or have been collected and made available by security experts, such as the `Xato` dataset. These datasets have been used to study the following problems: (1) do users avoid selecting known bad passwords and (2) do users really employ the rules suggested by information security policies and (3) do users exhibit a selection bias towards specific digits or special characters when creating a password.

To address the first issue, we examined the similarity, based on Levenshtein distance, between chronologically ordered datasets and a set of known bad passwords. Results were positive as we observed that there is a statistical significance in the dissimilarity among these files, suggesting that users are gradually improving in this direction, avoiding the selection of passwords similar to known bad passwords.

Regarding the second question, we applied machine learning clustering techniques on the datasets, in order to identify different clusters of users that select passwords that are more or less compliant with password selection standards. We have mapped passwords to a 10-dimensional space, in which each attribute quantifies the compliance against individual established policies, and then applied a $k$-means algorithm with Silhouette coefficient. We observed encouraging user practices such as the tendency to select longer passwords that make use of digits. However, users still omit most other security rules such as use of special characters, mixing lower with upper case letters and the use of publicly available information (names, emails etc.).

Lastly, we computed the frequency distributions of password length, special character and digit use in passwords, in all four datasets. Results reveal that the majority of users are indeed selecting passwords with lengths as those suggested by password policies. However, the frequency of certain digits and special characters appearing in the passwords reveal the existence of selection bias. SETA programs need to highlight and attempt to remedy the problems of selection bias and use of publicly available information.

Further study of this area is needed in order to investigate existing datasets and to look deeper into the recent password policies put forth by key industry players. In general, we would like to analyze newer datasets as they become available in order to corroborate if existing trends (e.g., increasing dissimilarity from bad passwords) persist and improve over time. We would also like to study the existence and nature of digit substitution patterns (e.g., *leet speak*) to identify password creation heuristics that ought to be avoided. Finally, of great interest is the frequency of (all) characters as they appear in specific positions within a password. The implication here is that knowing the observed frequency distribution per position could potentially help mitigate certain types of user authentication attacks.